\documentclass[12pt]{iopart} \usepackage{bm}
\usepackage{amsgen,amssymb,amsfonts,amsbsy}

\begin{document}

\comment[Radiation reaction in the 2.5PN waveform] {Radiation reaction in the
  2.5PN waveform from inspiralling binaries in circular orbits}

\author{Lawrence E. Kidder$^1$, Luc Blanchet$^2$, Bala R. Iyer$^3$}

\address{$^1$ Center for Radiophysics and Space Research, \\
Cornell University, Ithaca, New York, 14853  U.S.A.}
\address{$^2$ ${\mathcal{G}}{\mathbb{R}}\varepsilon{\mathbb{C}}{\mathcal{O}}$, 
Institut d'Astrophysique de Paris ---C.N.R.S., \\
98$^{\mathrm{bis}}$ boulevard Arago, 75014 Paris, France}
\address{$^3$ Raman Research Institute, Bangalore 560 080, India}

\eads{\mailto{kidder@astro.cornell.edu}, \mailto{blanchet@iap.fr}, 
\mailto{bri@rri.res.in}}

\begin{abstract}
In this Comment we compute the contributions of the radiation reaction force
in the 2.5 post-Newtonian (PN) gravitational wave polarizations for compact
binaries in circular orbits. (i) We point out and correct an inconsistency in
the derivation of~\cite{Arun2004}. (ii) We prove that all contributions from
radiation reaction in the 2.5PN waveform are actually negligible since they
can be absorbed into a modification of the orbital phase at the 5PN order.
\end{abstract}


\submitto{\CQG}

\section{Introduction}

The second and a half post-Newtonian (2.5PN$\sim c^{-5}$) waveform for
inspiralling compact binaries moving in quasi-circular orbits was computed by
Arun \textit{et al}~\cite{Arun2004}. Starting from the expressions of the
radiative multipole moments given by Eqs.~(3.4)--(3.6) of~\cite{Arun2004}, and
of the source moments given by~(3.16)--(3.18) of~\cite{Arun2004}, the wave
form is made of the instantaneous terms (Eqs.~(5.1)--(5.4)
of~\cite{Arun2004}), the hereditary memory-type contributions (Eqs.~(4.24)
of~\cite{Arun2004}), and the tail contributions (Eqs.~(4.38) corrected by the
published Erratum~\cite{Arun2004}). In this Comment we are concerned with the
piece of the instantaneous waveform that is the contribution from radiation
reaction (RR) at 2.5PN order in the dynamics of the inspiralling compact
binary. We point out and correct an inconsistency in the derivation of RR
terms in~\cite{Arun2004}, but argue that in fact the RR terms are negligible
in the 2.5PN waveform, since they can be absorbed into a 5PN order
contribution to the orbital phase evolution.

\section{Quasi-circular inspiral at 2.5PN order}

Adopting the conventions of~\cite{Arun2004}, $r = \vert{\bf x}\vert$ is the
binary separation with ${\bf x} = {\bf y}_1 - {\bf y}_2$ the vectorial
separation between the particles, ${\bf v} = {\bf v}_1 - {\bf v}_2$ is the
relative velocity, $m = m_1 + m_2$ is the total mass of the binary system
(distinct from the mass monopole moment $M$ given by the ADM mass of the
system), and $\nu = m_1 m_2/m^2$ is the reduced mass divided by the total
mass.

In modeling the orbital motion of the binary at 2.5PN order,
Ref.~\cite{Arun2004} stressed the importance of including the
radiation-reaction force in the 2.5PN expression of the binary
acceleration. Consider an arbitrary orbit confined to the x-y plane. The
relative position, velocity, and acceleration are given by
\begin{eqnarray}
{\bf x} = r {\bf n}, \\
\label{eq:vEOM}
{\bf v} = \dot r {\bf n} + r \omega {\bm \lambda}, \\
\label{eq:aEOM}
{\bf a} = (\ddot r - r \omega^2) {\bf n} + 
(r \dot \omega + 2 \dot r \omega) {\bm \lambda}.
\end{eqnarray}
Here ${\bm \lambda} = {\bf \hat z} \times {\bf n}$ where ${\bf \hat z}$ is the
unit vector along the z-direction. The orbital frequency $\omega$ is related
as usual to the orbital phase $\phi$ via $\omega = \dot \phi$.

Through 2PN order, it is possible to model the motion of the binary as a
circular orbit with the solution $\ddot r = \dot r = \dot \omega = 0$ and $r
\omega^2 = - {\bf n} \cdot {\bf a}$. Detailed calculations of the 2PN
equations of motion for a circular orbit (in harmonic coordinates) yield
\begin{equation}
\label{om2PN}
\omega^2 = \frac{Gm}{r^3} \left\{ 1 + \left(-3 + \nu \right) \frac{G m}{r c^2}
+ \left(6 + \frac{41}{4} \nu + \nu^2 \right) \left(\frac{G m}{r c^2}\right)^2
\right\}.
\end{equation}

At 2.5PN order, however, the effect of the inspiral motion must be taken into
account. The leading order contribution to the inspiral can be obtained by
examining the Newtonian orbital energy,
\begin{equation}
\label{Eorbital}
E = - \frac{1}{2} \frac{Gm^2\nu}{r},
\end{equation}
and the leading order gravitational-wave luminosity,
\begin{equation}
\label{luminosity}
- \frac{d E}{dt} = \frac{32}{5} \frac{G^4 m^5\nu^2}{r^5 c^5}.
\end{equation}
Here we assume that the energy radiated by the gravitational waves is balanced
by the change in the orbital energy. This yields
\begin{equation}
\label{eq:rDot}
\dot r = \frac{dE/dt}{dE/dr} = - \frac{64}{5} \frac{G^3 m^3\nu}{r^3 c^5}.
\end{equation}
Similarly the orbital frequency changes by (using $G m = r^3 \omega^2$ at
leading order)
\begin{equation}
\label{eq:OmegaDot}
\dot \omega = \frac{d E/dt}{d E/d\omega} = \frac{96}{5} \frac{G
m\nu}{r^3}\left(\frac{G m}{r c^2}\right)^{5/2},
\end{equation}
while the orbital phase $\phi=\int\omega dt$, found by
integrating~(\ref{eq:OmegaDot}), is
\begin{equation}
\label{eq:phase}
\phi = - \frac{1}{32 \nu} \left(\frac{G m}{r c^2}\right)^{-5/2}.
\end{equation}
Substituting~(\ref{eq:rDot}) and~(\ref{eq:OmegaDot})
into~(\ref{eq:vEOM})--(\ref{eq:aEOM}), and noting that $\ddot r\sim
\mathcal{O}(c^{-10})$ is of the order of the \textit{square} of RR effects,
the expressions for the 2.5PN inspiral relative velocity and acceleration in
harmonic coordinates are obtained,
\begin{eqnarray}
\label{eq:V2.5PN}
{\bf v} &=& r \omega {\bm \lambda} - \frac{64}{5} \frac{G^3 m^3\nu}{r^3 c^5}
{\bf n},\\
\label{eq:a2.5PN}
{\bf a} &=& -\omega^2 {\bf x} - \frac{32}{5}
\frac{G^3 m^3 \nu}{r^4 c^5} {\bf v},
\end{eqnarray}
where $\omega$ is given by~(\ref{om2PN}). As we shall detail the 2.5PN RR
terms in both the inspiral velocity~(\ref{eq:V2.5PN}) and
acceleration~(\ref{eq:a2.5PN}) should be substituted into the gravitational
waveform at 2.5PN order.

\section{Computation of polarization waveforms}

In the leading quadrupole approximation the gravitational waveform is given by
\begin{equation}
\label{eq:QuadrupoleWaveform}
h_{ij}^{\mathrm{TT}} = \frac{2G}{c^4 R} {\cal P}_{ijkl}({\bf N}) \Bigl[
\ddot{I}_{kl} + \mathcal{O}(c^{-1})\Bigr],
\end{equation}
where $R$ is the distance to the observer and ${\cal P}_{ijkl}$ is the
transverse-traceless (TT) projection operator ${\cal P}_{ijkl}={\cal
P}_{ik}{\cal P}_{jl}-\frac{1}{2}{\cal P}_{ij}{\cal P}_{kl}$, with ${\cal
P}_{ij}=\delta_{ij}-N_iN_j$ and ${\bf N}=(N_i)$ the radial direction from the
source to the observer. Here $\ddot{I}_{kl}$ is the second time-derivative of
some appropriate source quadrupole moment defined from a general
post-Newtonian multipole moment formalism. The remainder $\mathcal{O}(c^{-1})$
indicates higher PN corrections coming notably from all the higher multipolar
orders.

Defining an orthonormal triad $({\bf N},{\bf P},{\bf Q})$ where ${\bf P}$ and
${\bf Q}$ are unit polarization vectors transverse to the direction of
propagation, the polarization waveforms $h_+$ and $h_\times$ are computed from
the waveform~(\ref{eq:QuadrupoleWaveform}) by
\begin{eqnarray}
\label{eq:PlusPolarization}
h_+ &=& \hbox{$\frac{1}{2}$} ( P_i P_j - Q_i Q_j ) h_{ij}^{\mathrm{TT}}, \\
\label{eq:CrossPolarization}
h_\times &=& \hbox{$\frac{1}{2}$} ( P_i Q_j + P_j Q_i ) h_{ij}^{\mathrm{TT}}.
\end{eqnarray}
If the orbital plane is chosen to be the x-y plane with the orbital phase
$\phi$ measuring the direction of the unit vector ${\bf n} = {\bf x}/r$ along
the relative separation vector, then
\begin{equation}
{\bf n} = \cos{\phi} {\bf \hat x} + \sin{\phi} {\bf \hat y}.
\end{equation}
Ref.~\cite{Arun2004} has chosen the polarization vector {\bf P} to lie along
the x-axis and the observer to be in the y-z plane with
\begin{equation}
{\bf N} = \sin{i} {\bf \hat y} + \cos{i} {\bf \hat z},
\end{equation}
where $i$ is the orbit's inclination angle ($0\leq i\leq\pi$). With these
definitions ${\bf P}$ lies along the intersection of the orbital plane with
the plane of the sky in the direction of the ascending node, and the orbital
phase $\phi$ is the angle between the ascending node and the direction of body
one (say). The rotating orthonormal triad $({\bf n},{\bm \lambda},{\bf \hat
z})$ describing the motion of the binary is then related to the polarization
triad $({\bf N},{\bf P},{\bf Q})$ by
\begin{eqnarray}
\label{eq:SourceToObserver}
{\bf n} &=& 
\cos{\phi} {\bf P} + \sin{\phi} (c_i {\bf Q} + s_i {\bf N}), \\
{\bm \lambda} &=& 
- \sin{\phi} {\bf P} + \cos{\phi} (c_i {\bf Q} + s_i {\bf N}), \\
{\bf \hat z} &=& - s_i {\bf Q} + c_i {\bf N},
\end{eqnarray}
where we pose $c_i=\cos{i}$ and $s_i=\sin{i}$.

\section{Radiation reaction contributions to the waveform}

All contributions arising from the gravitational RR are contained in the
leading order quadrupolar waveform given
by~(\ref{eq:QuadrupoleWaveform}). They have three different origins. The first
type of RR term is issued directly from the expression of the (post-Newtonian)
quadrupole moment $I_{ij}$ at order 2.5PN, and reads (see Eq.~(3.16a)
of~\cite{Arun2004})
\begin{equation}
\label{Iij}
I_{ij} = \nu m x^{<i}x^{j>} + \cdots + \frac{48}{7}\frac{G^2m^3\nu^2}{r
c^5}x^{<i}v^{j>}.
\end{equation}
The first term is the usual expression of the Newtonian quadrupole moment (the
brackets surrounding indices refer to the symmetric-trace-free
projection). The dots indicate the 1PN and 2PN conservative terms, and terms
strictly higher than 2.5PN. Taking two time-derivatives of~(\ref{Iij}) we get
\begin{equation}
\label{ddIij0}
\ddot{I}_{ij} = 2 \nu m \bigl(v^{<i}v^{j>} + x^{<i}a^{j>}\bigr) + \cdots -
\frac{192}{7}\frac{G^3m^4\nu^2}{r^4 c^5} x^{<i}v^{j>}.
\end{equation}

The second type of RR term comes from inserting the expression of the
acceleration ${\bf a}$ given by~(\ref{eq:a2.5PN}), which produces another RR
term which modifies the term in~(\ref{ddIij0}) as
\begin{equation}
\label{ddIij}
\ddot{I}_{ij} = 2 \nu m \Bigl(v^{<i}v^{j>} - \frac{G m}{r^3}
x^{<i}x^{j>}\Bigr) + \cdots - \frac{1408}{35}\frac{G^3m^4\nu^2}{r^4 c^5}
x^{<i}v^{j>}.
\end{equation} 
The sum of these RR contributions is exactly what has been computed
in~\cite{Arun2004}, where it was denoted by $\rho_{ij}^{(5/2)}$ and given by
Eq.~(5.2). We now replace~(\ref{ddIij}) into the
waveform~(\ref{eq:QuadrupoleWaveform}) and compute the two GW polarizations
according to~(\ref{eq:PlusPolarization})--(\ref{eq:CrossPolarization}), hence
\begin{equation}
\label{polar}
\hspace{-2.2cm}h_{+,\times} = \frac{2G}{c^4 R}
\left(\begin{array}{l}\frac{P_iP_j-Q_iQ_j}{2}\\[0.2cm]\frac{P_iQ_j+P_jQ_i}{2}
\end{array}\right)\Bigl[2
\nu m \Bigl(v^{i}v^{j} - \frac{G m}{r^3} x^{i}x^{j}\Bigr) + \cdots -
\frac{1408}{35}\frac{G^3m^4\nu^2}{r^4 c^5} x^{i}v^{j}\Bigr].
\end{equation}
For ease of notation, in Eq.~(\ref{polar}) above and similar equations
later the first row (line) corresponds to $+$ and the second row
(line) to $\times$. The RR term in~(\ref{ddIij}) has already been
included in the final result of~\cite{Arun2004}. Its contribution to
the polarization waveforms is given by
\begin{equation}
\label{delta12}
\delta_{12} h_{+,\times} = \frac{2G\nu m x}{c^2 R}
\left\{\begin{array}{l} \displaystyle \frac{352}{35} (1 + c_i^2) \nu x^{5/2}
\sin{2\phi},\\[0.5cm] \displaystyle - \frac{352}{35} (2 c_i) \nu x^{5/2}
\cos{2\phi},\end{array}\right.
\end{equation}
where we pose $x = (Gm\omega/c^3)^{2/3}$. With the notation $\delta_{12}$ we
remind that this term was made of two distincts pieces.

However, let us show there is also another contribution to RR that has been
overlooked in~\cite{Arun2004}. Indeed the instantaneous terms in the waveform
$h_{ij}^{\mathrm{TT}}$ (Eqs. (5.1)--(5.4) of~\cite{Arun2004}) are given in
terms of the relative position ${\bf n}$ and velocity ${\bf v}$ of the binary,
as well as the PN parameter $\gamma = G m/(r c^2)$. In computing the
polarization waveforms, \textit{i.e.} projecting out $h_{ij}^{\mathrm{TT}}$ to
get $h_{+,\times}$, Ref.~\cite{Arun2004} has substituted, at the last stage of
the computation, ${\bf v} = r \omega {\bm \lambda}$ in the waveform to obtain
Eqs.~(5.9)--(5.10) of~\cite{Arun2004}. This is correct for all the terms {\it
except} the two leading order ``Newtonian'' terms in~(\ref{polar}) for which
one must use the true expression of the velocity~(\ref{eq:V2.5PN}) {\it
including} RR inspiral. We find that substituting the inspiral
velocity~(\ref{eq:V2.5PN}) into the leading terms of the
waveform~(\ref{polar}) yields the following additional contribution to the
polarization waveforms,
\begin{equation}
\label{delta3}
\delta_3 h_{+,\times} = \frac{2G\nu m x}{c^2 R} \left\{\begin{array}{l}
\displaystyle \frac{64}{5} (1 + c_i^2) \nu x^{5/2} \sin{2\phi},\\[0.5cm]
\displaystyle - \frac{64}{5} (2 c_i) \nu x^{5/2}
\cos{2\phi}.\end{array}\right.
\end{equation}
Notice that both results~(\ref{delta12}) and~(\ref{delta3}) have the same
structure. Since the contribution~(\ref{delta3}) has not been taken into
account in~\cite{Arun2004} we find that the following terms in the 2.5PN
polarization waveforms of~\cite{Arun2004} (Eqs.~(5)--(6) of the Erratum) are
to be changed from
\begin{equation}
\label{Hold}
\hspace{-1.5cm}H_{+,\times}^{(2.5)}\vert_\mathrm{old} =
\left\{\begin{array}{l} \displaystyle \cdots + \sin{2\psi} \Bigl[ -\frac{9}{5}
+ \frac{14}{5} c_i^2 + \frac{7}{5} c_i^4 + \nu\Bigl(\frac{96}{5} - \frac{8}{5}
c_i^2 - \frac{28}{5} c_i^4\Bigr)\Bigr] + \cdots,\\[0.5cm] \displaystyle \cdots
+ c_i \cos{2 \psi} \Bigl[ 2 - \frac{22}{5} c_i^2 + \nu\Bigl(-\frac{154}{5} +
\frac{94}{5} c_i^2\Bigr)\Bigr] + \cdots,\end{array}\right.
\end{equation}
to
\begin{equation}
\label{Hnew}
\hspace{-1.5cm}H_{+,\times}^{(2.5)} = \left\{\begin{array}{l} \displaystyle
\cdots + \sin{2\psi} \Bigl[ -\frac{9}{5} + \frac{14}{5} c_i^2 + \frac{7}{5}
c_i^4 + \nu\Bigl(32 + \frac{56}{5} c_i^2 - \frac{28}{5} c_i^4\Bigr)\Bigr] +
\cdots,\\[0.5cm] \displaystyle \cdots + c_i \cos{2 \psi} \Bigl[ 2 -
\frac{22}{5} c_i^2 + \nu\Bigl(-\frac{282}{5} + \frac{94}{5} c_i^2\Bigr)\Bigr]
+ \cdots.\end{array}\right.
\end{equation}
We recall that the phase variable $\psi$ differs from $\phi$ and is given by
Eq.~(5.6) of~\cite{Arun2004}. The difference between $\psi$ and $\phi$ is at
order 4PN at least (see~\cite{Arun2004} for discussion). Note that the new
correction~(\ref{delta3}) is in terms that vanish in the limit $\nu \to 0$,
and so is still consistent with the results of black hole perturbation
theory~\cite{Tagoshi1994}.

\section{Phase modulation due to radiation reaction}

We are now going to show that \textit{all} the RR contributions are in fact
negligible in the 2.5PN waveform. From the sum of the previous
results~(\ref{delta12}) and~(\ref{delta3}) we end up with the following total
contribution due to RR in the GW polarizations:
\begin{equation}
\label{RR}
\delta^{\mathrm{RR}} h_{+,\times} = \frac{2G\nu m x}{c^2 R}
\left\{\begin{array}{l} \displaystyle \frac{160}{7} (1 + c_i^2) \nu x^{5/2}
\sin{2\phi},\\[0.5cm] \displaystyle - \frac{160}{7} (2 c_i) \nu x^{5/2}
\cos{2\phi}.\end{array}\right.
\end{equation}
Comparing this result with the GW polarizations at Newtonian order,
\begin{equation}
\label{hN}
h_{+,\times}^{(\mathrm{N})} = \frac{2G\nu m x}{c^2 R} \left\{\begin{array}{l}
\displaystyle - (1 + c_i^2) \cos{2\phi},\\[0.5cm] \displaystyle - (2 c_i)
\sin{2\phi},\end{array}\right.
\end{equation}
we see that the RR terms can be absorbed into a redefinition of the phase
variable as
\begin{equation}
\label{Phi}
\Phi = \phi + \frac{80}{7} \nu x^{5/2}.
\end{equation}
This means that the 2.5PN waveform will take exactly the same expression as if
we had neglected the RR contributions~(\ref{RR}), but with phase variable
$\Phi$ in place of $\phi$. Now the point is that the added term in~(\ref{Phi})
represents an extremely small modification of the phasing. Indeed, we take
into account the expression for the phase as a function of frequency at the
leading order (\textit{i.e.} due to leading order RR effects), which has been
computed in~(\ref{eq:phase}) and whose order $\sim c^5$ is the
\textit{inverse} of the order $\sim c^{-5}$ of RR effects. Thus we find that
the redefined phase is equivalent to
\begin{equation}
\Phi = - \frac{x^{-5/2}}{32\nu}\Bigr[1 + \cdots - \frac{2560}{7} \nu^2 x^{5}
  \Bigr].
\end{equation}
This means that the RR terms seen as modulations of the phase evolution,
contribute to the phase much beyond the 2.5PN order, namely at order 5PN$\sim
x^5$ beyond the leading phase evolution. Indeed there is in principle no point
in including these terms because they are comparable to 5PN terms in the phase
that are unknown --- only the phase evolution up to 3.5PN order is known.

We conclude therefore that the RR terms can in fact be neglected in the 2.5PN
waveform for circular orbits. This is similar to what happens in the energy
flux for circular orbits where we know that the 2.5PN radiation reaction gives
finally a contribution only at 5PN order~\cite{B96} (the 2.5PN terms in the
energy flux are only due to tails). Hence we can by the redefinition of the
phase variable~(\ref{Phi}) completely remove the RR contribution given
by~(\ref{RR}). However we can also decide to include such terms in the 2.5PN
waveform (as usual there are many different ways of presenting PN results at a
given order of approximation). Here we simply propose to include the RR terms
as they are in the templates of binary inspiral, \textit{i.e.} by correcting
the inconsistency in~\cite{Arun2004} (which has by the above argument no
physical effect at 2.5PN) and keeping the corrected waveform in the form given
by Eqs.~(\ref{Hnew}). For completeness, we note that if we choose to include
all the RR terms into the phase redefinition $\Phi$, Eq.~(\ref{Hnew}) will
then be modified into
\begin{equation}
\label{Halt}
\hspace{-1.5cm}{H'}_{+,\times}^{(2.5)} = \left\{\begin{array}{l} \displaystyle
\cdots + \sin{2\Psi} \Bigl[ -\frac{9}{5} + \frac{14}{5} c_i^2 + \frac{7}{5}
c_i^4 + \nu\Bigl(\frac{64}{7} - \frac{408}{35} c_i^2 - \frac{28}{5}
c_i^4\Bigr)\Bigr] + \cdots,\\[0.5cm] \displaystyle \cdots + c_i \cos{2 \Psi}
\Bigl[ 2 - \frac{22}{5} c_i^2 + \nu\Bigl(-\frac{374}{35} + \frac{94}{5}
c_i^2\Bigr)\Bigr] + \cdots,\end{array}\right.
\end{equation}
where $\Psi$ is equal to $\psi$ plus the added RR contribution given by the
second term in~(\ref{Phi}), and where $\psi$ itself differs from $\phi$ by
Eq.~(5.6) in~\cite{Arun2004}. (All the other terms in the waveform are then to
be expressed with the same phase variable $\Psi$.)

\ack This work was supported in part by grants from the Sherman Fairchild
Foundation, NSF grants PHY-0354631, DMS-0553677 and NASA grant NNG05GG51G at
Cornell. LB and BRI thank the Indo-French Collaboration (IFCPAR) for its
support.

\end{document}